# Offset-locking-based frequency stabilization of external cavity diode lasers for long-distance quantum communication

Takuto Miyashita[1]*, Takeshi Kondo[1], Kohei Ikeda[1], Kazumichi Yoshii[1], Feng-Lei Hong[1], and Tomoyuki Horikiri[1,2]*

[1]*Department of Physics, Graduate School of Engineering Science, Yokohama National University, Yokohama, Kanagawa 240-8501, Japan*
[2]*JST PRESTO, Kawaguchi, Saitama 332-0012, Japan*

E-mail: takuto_miyashita@outlook.com, horikiri-tomoyuki-bh@ynu.ac.jp

Quantum repeaters are required for long-distance quantum communication. For efficient coupling of quantum entangled photon sources with narrow-linewidth quantum memories we performed the frequency stabilization of two lasers at 1514 and 1010 nm. The 1514 nm pump laser of the entangled photon source exhibited a frequency stability of $3.6 \times 10^{-12}$ ($\tau = 1$ s). The 1010 nm pump laser of the wavelength conversion system exhibited a frequency stability of $3.4 \times 10^{-12}$ ($\tau = 1$ s). The stabilities of both lasers were approximately two orders of magnitude smaller than the frequency width of 4 MHz of the Pr:YSO quantum memory. Such frequency-stabilized lasers can realize the remote coupling of a quantum memory and an entangled photon source in quantum repeaters.

## 1. Introduction

Frequency-stabilized lasers are indispensable tools in various domains including metrology, quantum optics, and quantum information science. In the field of quantum information science, to address the increasing expectations of unconditional security in telecommunications, quantum communication technology has been developed. Quantum communication is expected to realize not only secure communication but also quantum internet, which can connect quantum devices located worldwide.[1]

Existing optical fiber-based quantum communication involves a distance limit of several hundred kilometers owing to the photon transmission loss of the optical fiber (~0.2 dB/km at telecommunication wavelength). To realize quantum communication exceeding the distance limit, a scheme involving quantum repeaters[2] must be adopted in which the distance between the repeater nodes is tens of kilometers or less. The total communication distance (distance between end nodes) can be increased by entanglement swapping and purification.[2]

In typical quantum repeater schematics, quantum memories are installed in quantum repeater nodes to store and recall the quantum states of photons transmitted through optical fibers. Among various proposed and developed quantum memories, quantum memories that can realize multiplexed quantum communication is desirable to extend the distance and achieve a high throughput. The target quantum memory utilizes the atomic frequency comb (AFC)[3] formed inside rare earth ion-doped crystals. In this method, comb-shaped absorption lines are created in a wide inhomogeneous absorption spectrum. The incoming photon is absorbed by the AFC (to preserve the quantum state of this photon) and emitted after a certain period (to recall the quantum state of the incoming photon) to implement the repeater procedure including entanglement swapping and purification.

The key features of AFC include high multiplicity (>100)[4] as a quantum memory, high absorption/reproduction efficiency close to unity,[5] and considerable storage time (milliseconds or longer).[6] Several rare earth ion-doped materials are potential candidates to develop AFC quantum memories. The target material considered in this work is Pr:YSO,[7] which exhibits a relatively long coherence time even without the application of a large magnetic field, high efficiency, and multimodality. However, the quantum memory absorption wavelength of Pr:YSO is approximately 606 nm, which is considerably smaller than the communication wavelength of 1.5 μm used for long-distance communication.

Figure 1 shows the proposed quantum repeater scheme. A two-photon entangled source with a wide frequency band (>10 GHz) in the communication wavelength is often used as the entangled photon pair source. Entanglement swapping can be realized by storing the quantum states of two photons transmitted from adjacent two-photon sources in separate quantum memories (located in the same node) and recalling these states after both the photons arrive at the node. The distance of the shared entanglement can be increased after the entanglement swapping. For example, the initial entanglement distance, which is around 100 km in the scheme shown in Fig. 1, can be extended to 200 km by implementing entanglement swapping at the quantum repeater node. Moreover, the entanglement sharing is directly connected to the capability of quantum communication, as the shared entanglement can be used, for example, to realize the quantum state transfer and also quantum key distribution to achieve unconditionally secure classical communication. In this scheme, entangled photon sources generate two degenerate telecom photons. There is a scheme that uses non-degenerate sources instead of degenerate ones, but non-degenerate sources are generally more suitable for a scheme that place an entangled photon source in the repeater node where a quantum memory is installed.[8] This is because it can take advantage of the characteristics of the non-degenerate photon pair source, which has a quantum-memory-wavelength photon on one side and a telecommunication wavelength photon suitable for fiber transmission on the other side. In the present midpoint source scheme,[9,10] the entanglement distribution rate can be more than one order of magnitude higher when AFC memory is employed compared with the meet-in-the-middle scheme which uses nondegenerate photon pair generation.[11] In the case of the midpoint source, both entangled photon pairs are transmitted through optical fiber, which is an advantage of a degenerate telecommunication wavelength source. Therefore, in this study, we adopt a degenerate entangled photon source. The AFC quantum memory with Pr:YSO generally exhibits a narrow transition frequency width (<10 MHz), and thus, a narrow linewidth must be implemented in the entangled photon source to obtain a high bandwidth matching.

For the proposed quantum repeater scheme, we generated a narrow linewidth (~1 MHz) telecom two-photon source by using a cavity enhanced parametric downconversion (PDC) process.[12] Second harmonic generation (~750 nm) of a telecom laser is utilized as a pump laser for our degenerate PDC process. The telecom laser is also used for locking the cavity by Pound-Drever-Hall method. Therefore, the frequency stability of the generated telecom photons is guaranteed. Through this approach, the telecommunication wavelength entangled photons were expected to be coupled to the AFC quantum memories (the spectral domain in which the AFC is formed has a width of approximately 4 MHz in the case of Pr:YSO), without loss due to spectral mismatch. However, there is a difference in the wavelengths of the two-photon source and the quantum memory (approximately 606 nm in the case of Pr:YSO). To fill this gap, wavelength conversion must be performed. Sum frequency generation (SFG) of 1514 nm (198 THz) + 1010 nm (297 THz) → 606 nm (495 THz) is adopted to realize this conversion. The SFG process involves a telecommunication wavelength laser of 1514 nm, which is frequency doubled and used as the pump laser of the telecommunication wavelength two-photon, and a 1010 nm laser, which is the pump laser for the SFG wavelength conversion. Because it is necessary to absorb photons after wavelength conversion within the frequency width of approximately 4 MHz in AFC memory, the frequency stabilities of the two abovementioned lasers must be ensured. Therefore, the objective of this study is to stabilize the lasers to below the frequency width of the AFC memory. As the network grows, the relative frequency drift between all the sources will limit the scalability of the network. For this reason, we assume that the stability should be an order of magnitude smaller than the transition frequency width of 4 MHz. Furthermore, it is important to match the frequency of the created AFC with that of the photons to be stored.

Considering these aspects, we perform the frequency stabilization of two lasers, a telecommunication wavelength laser and a pump laser (1010 nm) for the SFG wavelength conversion. An optical frequency comb[13,14] which is phase-locked to the GPS signal (GPS-OFC) located in a quantum repeater node and an optical link[15] are used to stabilize the remote lasers. The adopted locking technique is the offset lock[16,17] realized using an electrical delay line to perform stabilization based on the GPS-OFC. In this manner, the frequency stability necessary to ensure bandwidth matching to an AFC quantum memory is realized. Also, the absolute frequency after the wavelength conversion can be adjusted to the absolute frequency of the AFC frequency. This convenient aspect has an advantage compared with the existing frequency stabilization methods that adopt atomic and/or molecular absorption lines. Furthermore, we compare the achieved laser stability with those of a stabilization technique involving the absorption line of molecular iodine.

## 2. Experiment

Figure 2 shows a schematic of the experimental setup. The pump lasers at 1514 nm and 1010 nm of the telecom two-photon and wavelength conversion, respectively, were stabilized to the GPS-OFC ($f_{rep}$ = 107 MHz) through offset locking with electrical delay lines. Both the lasers were external cavity diode lasers (ECDLs) (Sacher Serval Plus for 1514 nm light, and Toptica TA pro for 1010 nm) with free-running linewidths of approximately 300 kHz and 40 kHz, respectively, and a maximum output power of approximately 1 W.

The frequency stabilization setups of the two ECDLs were identical. Immediately after each laser was emitted, a certain amount of power was supplied to the wavelength conversion system through a beam splitter. The light used for stabilization was mixed with the output of the GPS-OFC by using a 50:50 branch coupler, and the beat signal was generated using a photo detector (PD). The output power of the beat signal was split by a directional coupler and sent to a frequency counter for frequency measurement. The beat signal passed through a 30 MHz bandpass filter and an amplifier, and an error signal was created using a delay-line offset locking. The length of the delay line was related to the offset frequency of the frequency lock. In this study, a coaxial cable with a length of 5 m was used, and the offset frequency was set as 30 MHz. The generated error signal was input to the servo controller (Laselock) and fed back to the laser current to realize the zero-crossing point of the error signal. The delay line and a mixer for error signal generation were covered with a heat insulating foamed styrol to reduce the temperature fluctuation.[18] Specifically, the temperature was maintained constant to ensure that the signal delay time did not change with the change of the room temperature.

## 3. Experimental results

Figures 3 and 4 show the experimental results. Using a frequency counter, the frequency fluctuation of the beat signal used for the error signal generation was measured for one hour. Allan standard deviations were calculated, and the frequency stability of each laser was evaluated. The gate time of the counter was one second.

Figure 3 shows the results of the 1514 nm laser. As shown in Fig. 3(a), the frequency fluctuation range is 25 MHz per hour in the free running case. After the stabilization, the fluctuation range is suppressed to 25 kHz, approximately 1/1000 of the original value, as shown in Fig. 3(b). Figure 3(c) shows the Allan standard deviation calculated from the measurement results. The black dashed line indicates the Allan standard deviation corresponding to the frequency stability of 100 kHz, which is an order of magnitude smaller than the frequency width of the AFC quantum memory. The blue dots are calculated the Allan standard deviation using the measured beat frequency shown in Fig. 3(b). The values are less than or equivalent to the stabilization reference GPS-OFC (shown as gray dots in Fig. 3(c)) up to an average time $\tau$ of 3 s. Since the measured beat signal is an in-loop signal in the servo, it means that the offset-locked 1514 nm laser follows the stabilization reference GPS-OFC well and the stability is limited to that of the GPS-OFC. In contrast, when $\tau > 3$ s, the stability of the in-loop beat becomes worse than that of the GPS-OFC. This means that when $\tau > 3$ s, the stability of the offset-locked 1514 nm laser is limited by the offset locking stability. To confirm the actual stability of the offset-locked laser, we also performed an out-of-loop signal measurement (red dashed line in Fig. 3(c)) using another OFC which is phase-locked to an iodine-stabilized Nd:YAG laser (narrow-linewidth OFC)[19] and has a much higher stability than that of the GPS-OFC. The stability of the out-of-loop signal indicates the stability of the offset-locked 1514 nm laser and is $3.6 \times 10^{-12}$ (corresponding to an absolute frequency of 0.71 kHz) at $\tau = 1$ s and $2.7 \times 10^{-11}$ (5.4 kHz) at $\tau = 287$ s, respectively. When $\tau > 300$ s, the Allan standard deviation of the offset locking 1514 nm laser decreases so that the target stability is maintained both in the short- and long-term, compared with the black dashed line.

Figure 4 shows the results of the 1010 nm laser, obtained in the same manner as those of the 1514 nm laser. The fluctuation range of 12 MHz per hour for the free-running case (Fig. 4(a)) is reduced to 10 kHz when locked, as shown in Fig. 4(b). As shown in Fig. 4(c), the Allan standard deviation of the in-loop signal (blue dots) is $1.7 \times 10^{-12}$ (corresponding to an absolute frequency of 0.49 kHz) at $\tau = 1$ s and $4.3 \times 10^{-12}$ (1.3 kHz) at $\tau = 393$ s, respectively. The stability of the in-loop signal is higher than that of the stabilization reference GPS-OFC across all the average times. This means that the offset-locked 1010 nm laser follows the stabilization reference GPS-OFC well and the stability is limited to that of the GPS-OFC. The Allan standard deviation of the GPS-OFC (gray dots) is $3.4 \times 10^{-12}$ (corresponding to an absolute frequency of 1.0 kHz) at $\tau = 1$ s and $7.2 \times 10^{-12}$ (2.1 kHz) at $\tau = 263$ s, respectively. Again, the stability of the offset-locked 1010 nm laser is better than the stability of 100 kHz (the black dashed line) both in the short- and long-term.

The free-running results shown in Figs. 3(a) and 4(a) highlight the performance difference of the 1514 nm and 1010 nm lasers, respectively, even under the same laboratory environment. The free-running linewidth specifications (300 kHz and 40 kHz for the 1514 nm and 1010 nm, respectively) are also consistent with the results. The better Allan standard deviation of the offset lock achieved for the 1010 nm laser compared with that of the 1514 nm laser is considered to be due to the better free-running stability and smaller linewidth. If higher stability is required, the slope of the error signal can be enlarged by changing the length of the delay line.[18]

## 4. Discussion and conclusion

We realized the frequency stability required to ensure reliable bandwidth matching to the quantum memory by frequency locking the lasers used in the quantum repeater scheme. The achieved frequency stability of $3\text{-}4\times10^{-12}$ at $\tau = 1$ s is limited by the stability of the GPS reference used in the locking. With a better frequency reference, the frequency stability of the offset-locked lasers could reach $1\text{-}2\times10^{-12}$ at 1 s (the in-loop Allan standard deviation). In atomic physics experiments such as optical lattice clocks, offset locking of laser frequencies has been applied and a frequency stability of $2.2\times10^{-13}$ at 1 s has been achieved.[20] In another experiment, an offset-locked compact solid-state laser, which can be used for wavelength conversion for connecting a telecom-wavelength photon and a NV center quantum memory, shows a frequency stability of $8.2\times10^{-11}$ at 1 s.[18] In the offset-locking method, the short-term instability at 1 s is determined by both signal-to-noise ratio and slope of the error signal at the locking point. The slope of the error signal can be enlarged by increasing the length of the delay line. However, this will also cause a reduction of the frequency capture range. Therefore, there is a trade-off relation between the frequency stability and capture range, which is important for reliable long-term operation. In the present study, the achieved frequency stability is sufficient for the purpose of connecting a two-photon source and a quantum memory.

During the offset locking, long-term temperature drift may result in a frequency shift. The error signal generated in the present study is sensitive to the coaxial cable length change, which could be caused by the temperature changes in the laboratory. Although the temperature change per day in the laboratory is less than 2 °C, frequency drift up to 10 kHz may occur in the frequency stabilization. In this regard, it was necessary to introduce a heat insulation system that can address a small temperature change at the delay line installation location. Nevertheless, temperature drift may occur in other electric circuit parts. This aspect must be examined when higher frequency stability is required in further work.

Considering the optical link, this setup is a realistic scheme that can be implemented by sending a high-intensity telecom laser to a remote node at which an OFC is located. The absolute frequency of the lasers involved in the quantum repeater scheme needs to be precisely monitored and adjusted to match the absolute frequency of the AFC. Figure 5 shows the experimental setup for the bandwidth matching of the AFC quantum memory by using frequency-stabilized lasers. Two wavelength conversion devices are adopted: one for AFC quantum memory bandwidth matching and one for absolute frequency monitoring. On the absolute frequency monitoring side, the frequency difference between the light of the AFC preparation laser and SFG wavelength conversion photon of approximately 606 nm is measured using a frequency counter. On the other side, to realize the AFC quantum memory bandwidth matching, a double-pass acousto-optic modulator is used as a frequency shifter to enable the frequency adjustment of the wavelength conversion pump laser (1010 nm).

Lasers with higher stability than the current system based on the GPS-OFC might be required to prevent relative frequency drift between individual links. The frequency stability can be improved by changing the stabilization method for each laser, for example, by stabilizing the lasers to the absorption lines of acetylene and molecular iodine.[21-23] Furthermore, the third harmonic generation can be adopted for locking a telecom laser to molecular iodine.[24]

We succeeded in stabilizing the wavelength conversion pump laser at 1010 nm using frequency doubling and modulation transfer spectroscopy[21] of molecular iodine. In Fig. 4(c), the light blue dots indicate the Allan standard deviation of the iodine-stabilized 1010 nm laser, which is measured using the narrow-linewidth comb.[19] The frequency stability was $9.1 \times 10^{-13}$ (corresponding to an absolute frequency of 0.27 kHz) at $\tau = 1$ s and $6.8 \times 10^{-13}$ (0.20 kHz) at $\tau = 155$ s, respectively. Locking was achieved to the a2 hyperfine transition of the *P*(41)56-0 line, as illustrated in the inset of Fig. 4(c). The stability was higher than that achieved using the delay line method. At present, the frequency stability of the iodine-stabilized laser does not follow the $1/\sqrt{\tau}$ characteristic and needs to be further improved. However, as shown in the figure, the minimum Allan standard deviation of the stability is almost more than one order of magnitude smaller than that of the GPS-OFC. By using a laser with a higher stability, the multiplicity, which is an excellent feature of AFC quantum memories, can be fully exploited.

Overall, we demonstrated the offset locking of pump lasers for quantum repeater systems. The method using the GPS-OFC as a reference of multiple lasers can be applied in an actual repeater system requiring the remote frequency locking of quantum devices.

## Acknowledgments

We thank Shuhei Tamura and Kazuya Niizeki for their support. This work was supported by SECOM foundation, JST PRESTO (JPMJPR1769), JST START (ST292008BN) and Japan Society for the Promotion of Science (JSPS) KAKENHI (JP20H02652).

# Figure Captions

**Fig. 1.** Quantum repeater scheme. For long-distance transmission, two photons are generated in the telecommunication wavelength. Sum frequency generation is used to realize the wavelength conversion to achieve coupling with the quantum memory at the repeater node. The two-photon source laser and the pump laser of the wavelength conversion are frequency-locked to the optical frequency comb. Eventually, the other photons from the two-photon sources are stored in the quantum memory at distant repeater nodes, leading to entanglement sharing between them. An optical link can be used to stabilize the lasers through a single optical frequency comb. The two-photon source lasers generate entangled states by generating two photons through spontaneous parametric downconversion (SPDC). TPS: Two-photon source including the TPS laser and nonlinear medium for SPDC, AFC QMs: atomic frequency comb quantum memories.

**Fig. 2.** Schematic of the frequency locking of the external cavity diode lasers (ECDLs). Offset locking with a delay line was performed using an optical frequency comb as a frequency reference. The optical frequency comb was stabilized to the GPS reference signal. The beat signal of the optical frequency comb and the ECDL was used to generate an error signal by the delay line. The servo controller returned feedback to the laser current. PD: Photo detector, DBM: Double balanced mixer, LPF: Low-pass filter, BPF: Bandpass filter. The optical and electrical paths are shown as solid and dashed lines, respectively.

**Fig. 3.** Measurement results of the 1514 nm laser. (a) 1 h frequency fluctuation for the free-running and frequency-locked cases. (b) Enlarged frequency fluctuation of the frequency-locked case in (a). (c) Allan standard deviation calculated from the frequency fluctuation. The black dashed line indicates the stability of 100 kHz, which is an order of magnitude smaller than the frequency width of the AFC quantum memory. Gray dots indicate the reference GPS-OFC value. Blue dots indicate the stability of the in-loop beat signal. The red dashed line corresponds to the out-of-loop signal, measured with the narrow-linewidth comb.

**Fig. 4.** Measurement results of the 1010 nm laser. (a) 1 h frequency fluctuation for the free-running and frequency-locked cases. (b) Enlarged frequency fluctuation of the frequency-locked case in (a). (c) Allan standard deviation calculated from the frequency fluctuation. The black dashed line indicates the stability of 100 kHz, which is an order of magnitude smaller than the frequency width of the AFC quantum memory. Gray dots indicate the stability of the GPS-OFC, and blue dots indicate the stability of in-loop signal. Light blue dots indicate the result of the frequency fluctuation when the 1010 nm laser was locked on a hyperfine transition of molecular iodine. The inset shows the hyperfine transitions of molecular iodine used in the lock.

**Fig. 5.** Absolute frequency adjustment setup. The figure shows the system used to match the absolute frequency of the wavelength conversion photon with that of the preparation laser for creating the AFC. The RF frequency is shifted by the acousto-optic modulator (AOM) after the 1010 nm laser is emitted. SFG: Sum frequency generation, HWP: Half wave plate, PBS: Polarizing beam splitter.

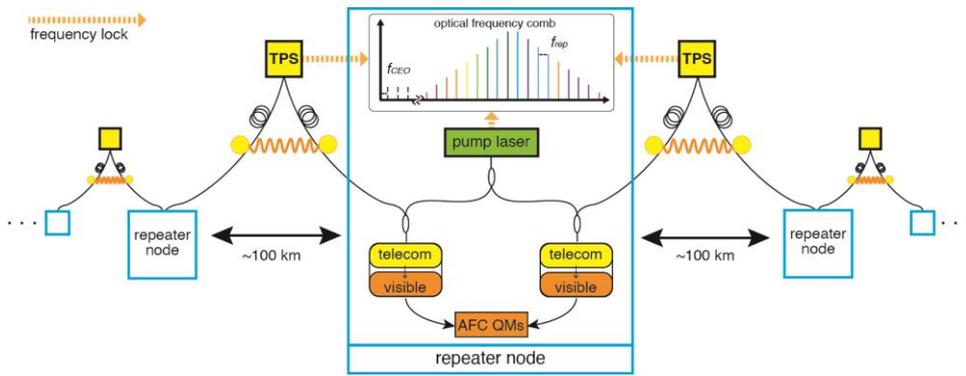

Fig.1.

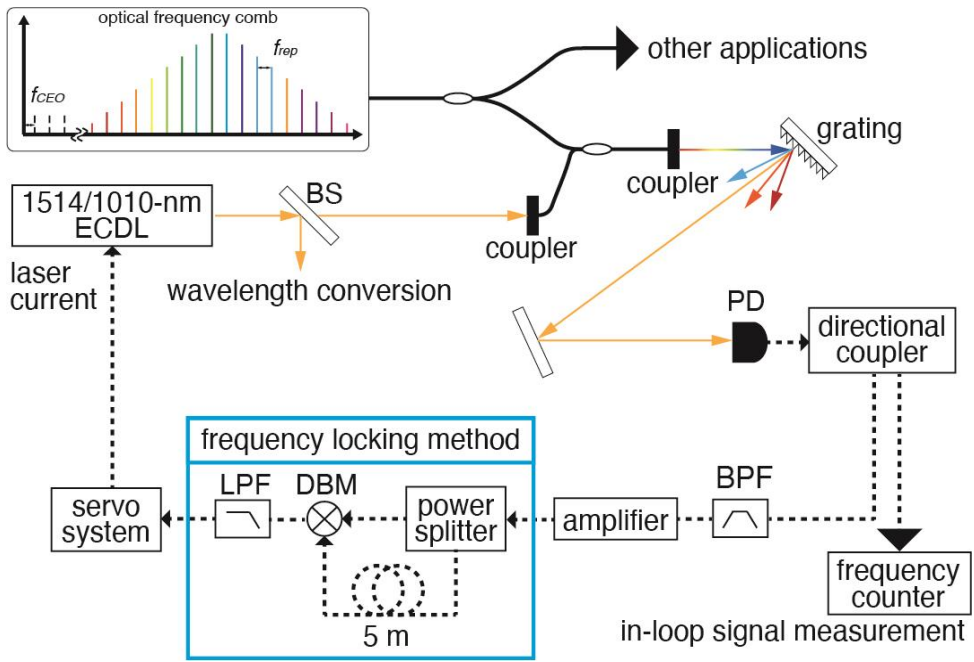

Fig.2.

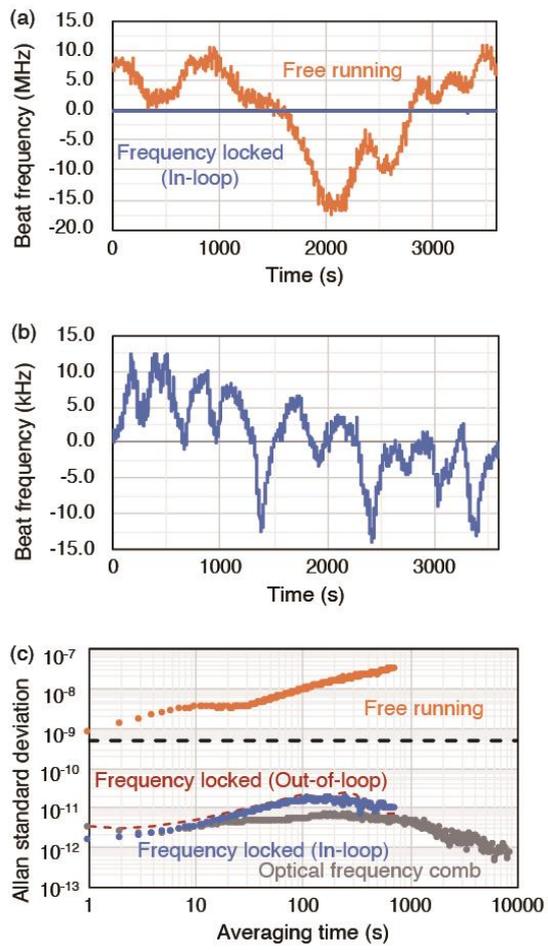

Fig.3.

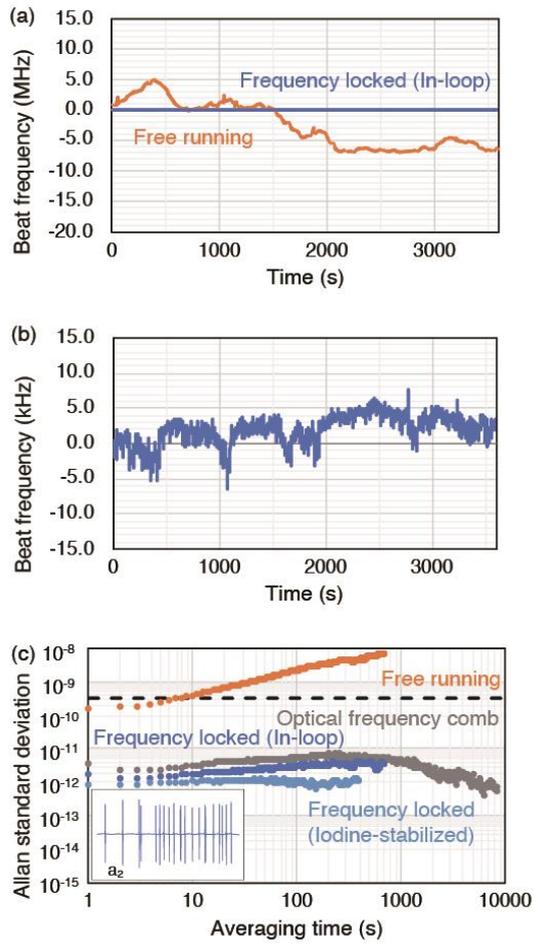

Fig.4.

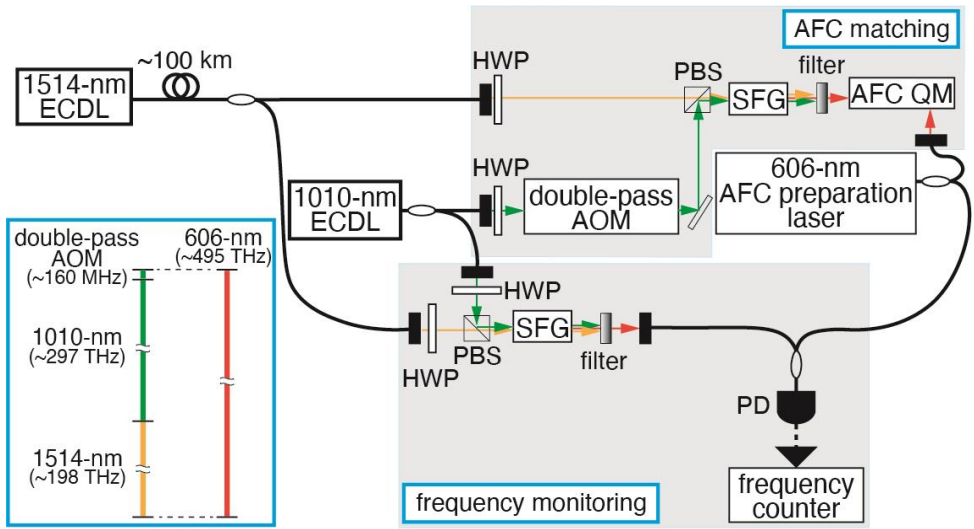

Fig. 5.